# Fitting Graphical Interaction Models to Multivariate Time Series

Michael Eichler
Department of Quantitative Economics
University of Maastricht
6200 MD Maastricht, Netherlands
m.eichler@ke.unimaas.nl

## Abstract

Graphical interaction models have become an important tool for analysing multivariate time series. In these models, the interrelationships among the components of a time series are described by undirected graphs in which the vertices depict the components while the edges indictate possible dependencies between the components. Current methods for the identification of the graphical structure are based on nonparametric spectral estimation, which prevents application of common model selection strategies. In this paper, we present a parametric approach for graphical interaction modelling of multivariate stationary time series. The proposed models generalize covariance selection models to the time series setting and are formulated in terms of inverse covariances. We show that these models correspond to vector autoregressive models under conditional independence constraints encoded by undirected graphs. Furthermore, we discuss maximum likelihood estimation based on Whittle's approximation to the log-likelihood function and propose an iterative method for solving the resulting likelihood equations. The concepts are illustrated by an example.

## 1 INTRODUCTION

Graphical models have become an important tool for analysing multivariate data. While the theory originally has been developed for variables that are sampled with independent replications, graphical models recently have been applied also to stationary multivariate time series (e.g., Brillinger 1996, Dahlhaus 2000, Eichler 1999, 2001, 2006, Dahlhaus and Eichler 2003).

A particularly simple graphical representation is provided by graphical interaction models, which visualize dependencies by undirected graphs. For time series, this approach has been discussed by Dahlhaus (2000), who introduced so-called partial correlation graphs. These are undirected graphs, in which each component of the time series is represented by one vertex. The conditional independences encoded by such graphs can be formulated in terms of the inverse of the spectral matrix. This has led to nonparametric tests for the presence of an edge in the partial correlation graph based on the maximum of the partial spectral coherence (Dahlhaus et al. 1997) or on the integrated partial spectral coherence (Eichler 2004). The concept of partial correlation graphs has been used in many applications from various scientific fields (e.g., Eichler et al. 2003, Timmer et al. 2000, Gather et al. 2002, Fried and Didelez 2003).

The main disadvantage of the current nonparametric approach to graphical modelling based on partial correlation graphs is the lack of a rigorous theory for identifying the best fitting graph. An alternative to the nonparametric approach is the fitting of parametric graphical models where the parameters are constrained with respect to undirected graphs. The problem of estimating the dependence structure of the process now becomes a problem of model selection where the best approximating model minimizes some chosen model distance such as the Kullback-Leibler information divergence.

In this paper, we propose graphical interaction models for stationary Gaussian time series that are defined in terms of inverse covariances. In these models, the conditional independences encoded by an undirected graph correspond to zero constraints on the parameters. In Section 2, we review the concept of conditional independence graphs for stationary time series. For Gaussian processes, these graphs are identical to partial correlation graphs, which are restricted to linear dependencies. In Section 3, we introduce para-metric graphical interaction models for Gaussian processes and discuss their relation to vector autoregressive models. In Section 4, we discuss estimation of the model parameters based on Whittle's approximation to the log-likelihood function; an iterative algorithm for computing the resulting maximum likelihood estimates is presented in Section 5. In Section 6, the fitting of graphical interaction models for time series is illustrated by an example with air pollution data, and Section 7 concludes.

## 2 UNDIRECTED GRAPHS FOR TIME SERIES

Let $\boldsymbol{X}_V = \big(\boldsymbol{X}_V(t)\big)_{t\in\mathbb{Z}}$ with $\boldsymbol{X}_V(t) = \big(X_v(t)\big)'_{v\in V}$ be a stationary Gaussian process with mean zero and covariances $\boldsymbol{\Gamma}_{VV}(u) = \mathbb{E}\boldsymbol{X}_V(t)\boldsymbol{X}_V(t-u)'$. Throughout the paper, we will make the following assumption.

**Assumption 2.1.** The spectral density matrix
$$\mathbf{f}_{VV}(\lambda) = \frac{1}{2\pi} \sum_{u=-\infty}^{\infty} \boldsymbol{\Gamma}_{VV}(u)\, e^{-\mathrm{i}\lambda u}$$
exists and its eigenvalues are bounded and bounded away from zero uniformly for all $\lambda \in [-\pi, \pi]$.

The structure of the interdependencies among the components of a multivariate time series $\boldsymbol{X}_V$ can be described in terms of conditional independencies between complete components of $\boldsymbol{X}_V$. More precisely, let $A$, $B$, and $S$ be disjoint subsets of $V$. Then the subprocesses $\boldsymbol{X}_A$ and $\boldsymbol{X}_B$ are said to be *conditionally independent* given $\boldsymbol{X}_S$ if
$$\mathbb{E}\big(g(\boldsymbol{X}_A)\,h(\boldsymbol{X}_B)|\boldsymbol{X}_S\big)$$
$$= \mathbb{E}\big(g(\boldsymbol{X}_A)|\boldsymbol{X}_S\big)\,\mathbb{E}\big(h(\boldsymbol{X}_B)|\boldsymbol{X}_S\big)$$
for all real-valued measurable functions $g(\cdot)$ and $h(\cdot)$ on $\mathbb{R}^{A\times\mathbb{Z}}$ and $\mathbb{R}^{B\times\mathbb{Z}}$, respectively. In this case, we write $\boldsymbol{X}_A \perp\!\!\!\perp \boldsymbol{X}_B \,|\, \boldsymbol{X}_S$.

As in the case of ordinary random variables, conditional independence relations of this form can be encoded by undirected graphs. Recall that an *undirected graph* $G$ is defined as a pair $(V,E)$ where $V$ is a set of *vertices* or *nodes* and $E$ is a collection of *undirected edges* $a \longrule b$ for distinct nodes $a,b \in V$. Then each component $\boldsymbol{X}_v$ of a process $\boldsymbol{X}_V$ may be represented by a single vertex $v$ in a graph $G = (V,E)$, while the edges in $E$ indicate possible dependencies among the components of $\boldsymbol{X}_V$. This leads to the following definition of *conditional independence graphs*.

**Definition 2.2.** Let $\boldsymbol{X}_V$ be a stationary process. The *conditional independence graph* associated with $\boldsymbol{X}_V$ is a graph $G = (V,E)$ with vertex set $V$ and edge set $E$ such that
$$a \longrule b \notin E \Leftrightarrow \boldsymbol{X}_a \perp\!\!\!\perp \boldsymbol{X}_b \,|\, \boldsymbol{X}_{V\setminus\{a,b\}}$$
for all distinct $a,b \in V$.

The conditional independence graph encodes the pairwise conditional independence relations that hold for the process $\boldsymbol{X}_V$. Under additional assumptions, more general conditional independence relations can be derived from it. Such properties that allow to associate certain statements about the graph $G$ with corresponding conditional independence statements about the variables in $\boldsymbol{X}_V$ are called graphical Markov properties. For instance, if $\boldsymbol{X}_V$ is a Gaussian process such that Assumption 2.1 holds, then $\boldsymbol{X}_V$ obeys the so-called global Markov property with respect to its conditional independence graph $G$: for disjoint sets $A, B, S \subseteq V$, we say that $S$ separates $A$ and $B$ if every path between $A$ and $B$ intersects $S$; this will be denoted as $A \bowtie B \,|\, S$. Then $\boldsymbol{X}_V$ satisfies the global Markov property with respect to $G$ if
$$A \bowtie B \,|\, S \text{ in } G \Rightarrow \boldsymbol{X}_A \perp\!\!\!\perp \boldsymbol{X}_B \,|\, \boldsymbol{X}_S$$
for all disjoint subsets $A, B, S \subseteq V$. We also say that $\boldsymbol{X}_V$ is Markov for the graph $G$. For details, we refer to Dahlhaus (2000).

For Gaussian processes $\boldsymbol{X}_V$, inference about conditional independence graphs is commonly based in the frequency domain (e.g., Dahlhaus 2000, Dahlhaus et al. 1997, Fried and Didelez 2003). Here, the conditional dependence between components $\boldsymbol{X}_A$ and $\boldsymbol{X}_B$ given $\boldsymbol{X}_S$ can be described by the partial cross-spectrum of $\boldsymbol{X}_A$ and $\boldsymbol{X}_B$ given $\boldsymbol{X}_S$,
$$\mathbf{f}_{AB|S}(\lambda) = \mathbf{f}_{AB}(\lambda) - \mathbf{f}_{AS}(\lambda)\mathbf{f}_{SS}(\lambda)^{-1}\mathbf{f}_{SB}(\lambda),$$
which is the cross-spectrum of the residual processes $\boldsymbol{\varepsilon}_{A|S}$ and $\boldsymbol{\varepsilon}_{B|S}$ obtained by removing the linear effects of the series $\boldsymbol{X}_S$ from the processes $\boldsymbol{X}_A$ and $\boldsymbol{X}_B$, respectively (cf Dahlhaus 2000). Equivalently, we can use the partial spectral coherency $\mathbf{R}_{AB|S}(\lambda)$ of $\boldsymbol{X}_A$ and $\boldsymbol{X}_B$ given $\boldsymbol{X}_S$, which is given by
$$R_{ab|S}(\lambda) = \frac{f_{ab|S}(\lambda)}{\sqrt{f_{aa|S}(\lambda)f_{bb|S}(\lambda)}}$$
for $a \in A$ and $b \in B$. For Gaussian processes, we have
$$\boldsymbol{X}_A \perp\!\!\!\perp \boldsymbol{X}_B \,|\, \boldsymbol{X}_S \Leftrightarrow \mathbf{R}_{AB|S}(\lambda) = 0 \ \ \forall \lambda \in [-\pi, \pi]. \quad (1)$$

For random vectors, the partial correlations can be obtained from the inverse of the covariance matrix. Dahlhaus (2000) showed that a similar relationship holds between the partial spectral coherences and the inverse of the spectral matrix. More precisely, let $\mathbf{g}_{VV}(\lambda) = \mathbf{f}_{VV}(\lambda)^{-1}$ denote the inverse spectral matrix. Then, under Assumption 2.1, we have
$$R_{ab|V\setminus\{a,b\}}(\lambda) = -\frac{g_{ab}(\lambda)}{\sqrt{g_{aa}(\lambda)g_{bb}(\lambda)}}. \quad (2)$$

This relation provides an efficient method for computing estimates for the partial spectral coherences $R_{ab|V\setminus\{a,b\}}(\lambda)$ of a process $\mathbf{X}_V$: let $\hat{\mathbf{f}}_{VV}(\lambda)$ be an estimate of the spectral matrix such as a nonparametric kernel spectral estimate and let $\hat{\mathbf{g}}_{VV}(\lambda) = \hat{\mathbf{f}}_{VV}(\lambda)^{-1}$ its inverse; then the partial spectral coherences can be estimated by substituting the entries of $\hat{\mathbf{g}}_{VV}(\lambda)$ for the corresponding entries of $\mathbf{g}_{VV}(\lambda)$ in (2). We note that for non-Gaussian processes, this nonparametric frequency domain approach can still be used for inference on the partial correlation graph, which describes only the linear dependencies among the variables.

## 3 PARAMETRIC MODELLING

The nonparametric frequency domain approach can easily be used for an exploratory analysis of the dependence structure of stationary time series, but it provides only insufficient tools for the identification of conditional independence graphs. Moreover, it does not include methods for building new, parsimonious models that can be used e.g. for forecasting. As an alternative, we consider parametric models that are Markov for a given graph $G$ and, thus, satisfy the conditional independencies encoded by $G$. Here, the main obstacle will be the constraints on the model parameters that are imposed by the graph: for classical time series models such as vector autoregressions, the constraints on the parameters are typically non-linear and prevent efficient algorithms for parameter estimation. Therefore we introduce graphical interaction models for stationary Gaussian processes, in which conditional independence restrictions correspond to zero constraints on the parameters.

### 3.1 VECTOR AUTOREGRESSIONS

To illustrate the problems encountered by imposing conditional independence restriction of the form $\mathbf{X}_a \perp\!\!\!\perp \mathbf{X}_b \,|\, \mathbf{X}_{V\setminus\{a,b\}}$ on a time series model, we briefly consider the case of vector autoregressive models. Suppose that $\mathbf{X}_V$ is a stationary process given by

$$\mathbf{X}_V(t) = \sum_{u=1}^{p} \mathbf{a}(u)\, \mathbf{X}_V(t-u) + \boldsymbol{\varepsilon}_V(t),$$

where $\boldsymbol{\varepsilon}_V$ is a Gaussian white noise process with mean zero and non-singular covariance matrix $\boldsymbol{\Sigma}$. Let $\mathbf{A}(z) = \mathbb{1} - \mathbf{a}(1)z - \ldots - \mathbf{a}(p)z^p$ be the characteristic polynomial of the process (e.g., Dahlhaus 2000), where $\mathbb{1}$ is the $V \times V$ identity matrix. Then, if $\det \mathbf{A}(z) \neq 0$ for all $z \in \mathbb{C}$ with $|z| \leq 1$, the inverse spectral matrix $\mathbf{f}(\lambda)^{-1}$ exists and is given by

$$\mathbf{f}(\lambda)^{-1} = 2\pi\, \mathbf{A}(e^{i\lambda})'\, \mathbf{K}\, \mathbf{A}(e^{-i\lambda}), \qquad (3)$$

where $\mathbf{K} = \boldsymbol{\Sigma}^{-1}$ is the inverse of the covariance matrix $\boldsymbol{\Sigma}$. From (1) and (2), it follows that $\mathbf{X}_a$ and $\mathbf{X}_b$ are conditionally independent given $\mathbf{X}_{V\setminus\{a,b\}}$ if and only if the corresponding entry in the inverse spectral matrix $\mathbf{f}(\lambda)^{-1}$ vanishes for all frequencies, that is,

$$\sum_{k,l=1}^{d} \sum_{u,v=0}^{p} K_{kl}\, a_{ka}(u)\, a_{lb}(v)\, e^{i\lambda(v-u)} = 0$$

for all $\lambda \in [-\pi, \pi]$, where $\mathbf{a}(0) = -\mathbb{1}$ and $\mathbf{a}(u) = 0$ whenever $u < 0$ or $u > p$. Thus, the relation $\mathbf{X}_a \perp\!\!\!\perp \mathbf{X}_b \,|\, \mathbf{X}_{V\setminus\{a,b\}}$ imposes the following $2p+1$ restrictions on the parameters:

$$\sum_{k,l=1}^{d} \sum_{u=0}^{p} K_{kl}\, a_{ka}(u)\, a_{lb}(u+h) = 0, \quad h = -p, \ldots, p.$$

It is clear from these expressions that estimation of the parameters, for example, by maximization of the likelihood function under such constraints would be difficult if not infeasible.

### 3.2 GRAPHICAL INTERACTION MODELS

Alternatively to the frequency domain approach, we can view the distribution of a Gaussian process $\mathbf{X}_V$ as a Gaussian Markov random field and describe the conditional independencies among the variables in $\mathbf{X}_V$ in terms of the inverse of the covariance matrix of $\mathbf{X}_V$. To make this idea explicit, assume that $\mathbf{X}_V$ satisfies Assumption 2.1. Then the inverse of the spectral matrix $\mathbf{f}(\lambda)$ exists and has an absolutely convergent Fourier expansion given by

$$\mathbf{f}(\lambda)^{-1} = 2\pi \sum_{u=-\infty}^{\infty} \boldsymbol{\Gamma}^{(i)}(u)\, e^{-i\lambda u}$$

(e.g., Bhansali 1980). Simple calculations show that the Fourier coefficients

$$\boldsymbol{\Gamma}^{(i)}(u) = \frac{1}{4\pi^2} \int_{\Pi} \mathbf{f}(\lambda)^{-1}\, e^{i\lambda u}, \quad u \in \mathbb{Z}, \qquad (4)$$

where $\Pi = [-\pi, \pi]$, satisfy

$$\sum_{u=-\infty}^{\infty} \boldsymbol{\Gamma}(v-u)\, \boldsymbol{\Gamma}^{(i)}(u) = \delta_{0u}\, \mathbb{1}$$

for all $v \in \mathbb{Z}$ (e.g., Shaman 1975, 1976). Thus, the matrix $\boldsymbol{\Gamma}^{(i)} = \left(\boldsymbol{\Gamma}^{(i)}(u-v)\right)_{u,v \in \mathbb{Z}}$ is the inverse of the covariance matrix $\boldsymbol{\Gamma}$ of the process $\mathbf{X}_V$, and we call $\boldsymbol{\Gamma}^{(i)}(u)$ the inverse covariances of $\mathbf{X}_V$.

Similarly as for ordinary Gaussian Markov random fields, the pairwise conditional independencies that hold for the process $\mathbf{X}_V$ can be expressed by zero entries in the inverse $\boldsymbol{\Gamma}^{(i)}$. In particular, (4) together with (1) and (2) leads to the following time domain characterization of conditional independencies between the components of $\mathbf{X}_V$.

**Proposition 3.1.** Let $\boldsymbol{X}_V$ be a stationary Gaussian process satisfying Assumption 2.1. Then, for all distinct $a, b \in V$,

$$\boldsymbol{X}_a \perp\!\!\!\perp \boldsymbol{X}_b \,|\, \boldsymbol{X}_{V \setminus \{a,b\}} \Leftrightarrow \Gamma^{(i)}_{ab}(u) = 0 \quad \forall u \in \mathbb{Z}.$$

The proposition suggests to define graphical interaction models directly in terms of inverse covariances and thus to make use of the zero constraints that are imposed on the parameters by the conditional independencies encoded by the absence of edges in the associated graph.

**Definition 3.2.** Let $\boldsymbol{X}_V$ be a stationary Gaussian process such that Assumption 2.1 holds. Furthermore, let $\boldsymbol{\Gamma}^{(i)}(u)$, $u \in \mathbb{Z}$, be the inverse covariances of $\boldsymbol{X}_V$. Then we say that $\boldsymbol{X}_V$ belongs to the *graphical interaction model* of order $p$ associated with the undirected graph $G = (V, E)$ if

$$\boldsymbol{\Gamma}^{(i)}(u) = 0 \quad \forall |u| > p \tag{5}$$

and for all distinct $a, b \in V$

$$a \text{---} b \notin E \Rightarrow \Gamma^{(i)}_{ab}(u) = \Gamma^{(i)}_{ba}(u) = 0 \quad \forall u \in \mathbb{Z}. \tag{6}$$

We will also say that the process $\boldsymbol{X}_V$ belongs to the GI($p$,$G$) model.

The graphical interaction model of order $p$ is parametrized by the vector of inverse covariances

$$\boldsymbol{\theta} = \begin{pmatrix} \operatorname{vech} \boldsymbol{\Gamma}^{(i)}(0) \\ \operatorname{vec} \boldsymbol{\Gamma}^{(i)}(1) \\ \vdots \\ \operatorname{vec} \boldsymbol{\Gamma}^{(i)}(p) \end{pmatrix},$$

where as usual the vec operator stacks the columns of a matrix and the vech operator stacks only the elements contained in the lower triangular submatrix. In the following, we denote the spectral matrices, covariances, and inverse covariances specified by the parameter $\boldsymbol{\theta}$ by $\mathbf{f}_{\boldsymbol{\theta}}(\lambda)$, $\mathbf{R}_{\boldsymbol{\theta}}(u)$, and $\boldsymbol{\Gamma}^{(i)}_{\boldsymbol{\theta}}(u)$, respectively. Then the parameter space $\boldsymbol{\Theta}(p, G)$ of the graphical interaction model of order $p$ associated with graph $G$ is the set of all $\boldsymbol{\theta} \in \mathbb{R}^r$ with $r = d^2 p + \frac{1}{2}d(d+1)$ such that

○ $\mathbf{f}_{\boldsymbol{\theta}}(\lambda)$ satisfies Assumption 2.1 and
○ $\boldsymbol{\Gamma}^{(i)}_{\boldsymbol{\theta}}(u)$ satisfy conditions (5) and (6).

We note that the latter condition imposes zero constraints on the elements of the parameter vector $\boldsymbol{\theta}$.

### 3.3 RELATION TO VECTOR AUTOREGRESSIONS

We now return to our discussion of vector autoregressive models. For this, suppose again that $\boldsymbol{X}_V$ is a stationary Gaussian VAR($p$) process that satisfies Assumption 2.1. Then equations (4) and (3) imply that the inverse covariance function $\boldsymbol{\Gamma}^{(i)}(u)$ of the process $\boldsymbol{X}_V$ is given by

$$\boldsymbol{\Gamma}^{(i)}(u) = \sum_{v=0}^{p-u} \mathbf{a}(v)' \, \mathbf{K} \, \mathbf{a}(v+u) \tag{7}$$

with $\mathbf{K} = \boldsymbol{\Sigma}^{-1}$. These equations show that the constraints on the autoregressive parameters $\mathbf{a}(u)$, $u = 1, \ldots, p$ and $\boldsymbol{\Sigma}$ can be reformulated in terms of the inverse covariances. More importantly, the equations also imply that the inverse covariances $\boldsymbol{\Gamma}^{(i)}(u)$ of a VAR($p$) process vanish for all $|u| > p$ (e.g., Bhansali 1980, Battaglia 1984). In other words, the equations show that every VAR($p$) process that is Markov for an undirected graph $G$ also belongs to the GI($p$,$G$) model.

Conversely, it can be shown that the equation system (7) has a unique solution of the autoregressive parameters $\mathbf{a}(u)$, $u = 1, \ldots, p$ and $\boldsymbol{\Sigma}$ in terms of the inverse covariances (Tunnicliffe Wilson 1972). Thus, a process $\boldsymbol{X}_V$ belongs to the GI($p$,$G$) model if and only if it belongs to the class of all VAR($p$) processes that are Markov for the graph $G$. Therefore the *graphical vector autoregressive model* of order $p$ associated with a graph $G$, denoted by VAR($p$,$G$), is identical to the GI($p$,$G$) model. Thus, graphical interaction models can be seen as graphical VAR models with an alternative parameterization that is better suited for representing the constraints on the parameters imposed by the graph.

## 4 LIKELIHOOD INFERENCE

In this section, we discuss estimation of the parameter $\boldsymbol{\theta}$ based on maximization of an appropriate version of the Gaussian log-likelihood function.

### 4.1 WHITTLE'S LIKELIHOOD

Suppose that $\boldsymbol{X}_V(1), \ldots, \boldsymbol{X}_V(T)$ are observations from a stationary Gaussian process $\boldsymbol{X}_V$. In the following, we assume that $\boldsymbol{X}_V$ belongs to a GI($p$,$G$) model with unknown parameter $\boldsymbol{\theta}_0 \in \boldsymbol{\Theta}(p, G)$. For estimation of the parameter $\boldsymbol{\theta}_0$, we consider the likelihood function

$$L_T(\boldsymbol{\theta}) = (2\pi)^{T/2} \big( \det \boldsymbol{\Gamma}_{\boldsymbol{\theta},T} \big)^{-1/2}$$
$$\cdot \exp\big(-\tfrac{1}{2} \boldsymbol{X}'_{V,T} \, \boldsymbol{\Gamma}^{-1}_{\boldsymbol{\theta},T} \, \boldsymbol{X}_{V,T}\big),$$

where $\boldsymbol{X}_{V,T} = \operatorname{vec}(\boldsymbol{X}_V(1), \ldots, \boldsymbol{X}_V(T))$ and $\boldsymbol{\Gamma}_{\boldsymbol{\theta},T}$ is the $|V|p \times |V|p$ matrix with entries $\boldsymbol{\Gamma}_{\boldsymbol{\theta}}(u - v)$ for $u, v = 1, \ldots, T$ (e.g., Brockwell and Davis 1991, §8.6). In maximum likelihood estimation, the parameter $\boldsymbol{\theta}_0$ is estimated by the vector in $\boldsymbol{\Theta}(p, G)$ that maximizes the likelihood function or, equivalently, that minimizes the

$-1/T$ log-likelihood function

$$\ell_T(\boldsymbol{\theta}) \sim \frac{1}{2T} \log \det \boldsymbol{\Gamma}_{\boldsymbol{\theta},T} + \frac{1}{2T} \boldsymbol{X}'_{V,T} \boldsymbol{\Gamma}^{-1}_{\boldsymbol{\theta},T} \boldsymbol{X}_{V,T}, \quad (8)$$

where we have omitted an additive constant. The main problem in the application of the log-likelihood function $\ell_T(\boldsymbol{\theta})$ to graphical interaction models are the constraints on the parameters. Since $\ell_T(\boldsymbol{\theta})$ depends on the parameters only through the covariance matrix $\boldsymbol{\Gamma}_{\boldsymbol{\theta},T}$ and its inverse $\boldsymbol{\Gamma}^{-1}_{\boldsymbol{\theta},T}$, derivation of the likelihood equations under the constraints of a graphical interaction model would be infeasible.

A more favourable choice for fitting graphical interaction models is Whittle's approximation to the exact Gaussian log-likelihood function, which has been originally proposed by Whittle (1953, 1954). Whittle's likelihood is formulated in terms of the inverse spectral matrix and thus allows a direct treatment of the constraints in graphical interaction models. More precisely, the approximation is based on the fact that the matrix $\boldsymbol{\Gamma}^{-1}_{\boldsymbol{\theta},T}$ in (8) can be approximated by the corresponding submatrix of inverse covariance matrix

$$\begin{pmatrix} \boldsymbol{\Gamma}^{(i)}_{\boldsymbol{\theta}}(0) & \cdots & \boldsymbol{\Gamma}^{(i)}_{\boldsymbol{\theta}}(1-T) \\ \vdots & \ddots & \vdots \\ \boldsymbol{\Gamma}^{(i)}_{\boldsymbol{\theta}}(T-1) & \cdots & \boldsymbol{\Gamma}^{(i)}_{\boldsymbol{\theta}}(0) \end{pmatrix}$$

(cf Shaman 1975, 1976). Together with the Szegö identity (cf Grenander and Szegö 1958), this leads to Whittle's likelihood function

$$\ell^{\mathrm{w}}_T(\boldsymbol{\theta}) = \frac{1}{4\pi} \int_\Pi \Big( \log \det \mathbf{f}_{\boldsymbol{\theta}}(\lambda) + \mathrm{tr}\big[\mathbf{I}^{(T)}(\lambda) \mathbf{f}_{\boldsymbol{\theta}}(\lambda)^{-1}\big] \Big) d\lambda.$$

In this expression, $\mathbf{I}^{(T)}(\lambda)$ is the periodogram matrix with entries

$$I^{(T)}_{ab}(\lambda) = \big(2\pi H_{2,T}\big)^{-1} d^{(T)}_a(\lambda) d^{(T)}_b(-\lambda),$$

where

$$d^{(T)}_a(\lambda) = \sum_{t=1}^T X_a(t) \exp(-\mathrm{i}\lambda t)$$

is the discrete Fourier transform of the data $X_a(1),\ldots,X_a(T)$. Minimization of Whittle's likelihood function leads to the Whittle estimator

$$\hat{\boldsymbol{\theta}}_T = \operatorname*{argmin}_{\boldsymbol{\theta} \in \boldsymbol{\Theta}(p,G)} \ell^{\mathrm{w}}_T(\boldsymbol{\theta}).$$

We note that in practice a tapered version of the periodogram should be used as this improves the small sample properties of the resulting Whittle estimate considerably (e.g., Dahlhaus 1988).

### 4.2 LIKELIHOOD EQUATIONS

The likelihood equations are the estimating equations obtained by setting the first derivatives of the log-likelihood function $\ell^{\mathrm{w}}_T(\boldsymbol{\theta})$ with respect to $\boldsymbol{\theta}$ to zero.

Using matrix calculus (e.g., Harville 1997), we obtain the following first derivatives of $\ell^{\mathrm{w}}_T(\boldsymbol{\theta})$

$$\frac{\partial \ell^{\mathrm{w}}_T(\boldsymbol{\theta})}{\partial \theta_i} = \frac{1}{4\pi} \int_\Pi \mathrm{tr}\Big[\big(\mathbf{I}^{(T)}(\lambda) - \mathbf{f}_{\boldsymbol{\theta}}(\lambda)\big) \frac{\partial \mathbf{f}_{\boldsymbol{\theta}}(\lambda)^{-1}}{\partial \theta_i}\Big] d\lambda. \quad (9)$$

Since the inverse spectral matrix is linear in the parameters, we get an explicit formula for its derivatives. Let $\theta_k$ correspond to $\Gamma^{(i)}_{ab}(u)$. Then, for $i,j \in V$,

$$\frac{\partial f^{-1}_{ij,\boldsymbol{\theta}}(\lambda)}{\partial \theta_k} = \begin{cases} 2\pi \delta_{ia}\delta_{ja} & \text{if } a=b \text{ and } u=0 \\ 2\pi \big[\delta_{ia}\delta_{jb}e^{-\mathrm{i}\lambda u} \\ \quad + \delta_{ib}\delta_{ja}e^{\mathrm{i}\lambda u}\big] & \text{otherwise} \end{cases}.$$

Substituted into (9), we therefore get

$$\frac{\partial \ell^{\mathrm{w}}_T(\boldsymbol{\theta})}{\partial \theta_k} = \int_\Pi \big(I^{(T)}_{ab}(\lambda) - f_{ab,\boldsymbol{\theta}}(\lambda)\big) e^{\mathrm{i}\lambda u} d\lambda \quad (10)$$

for all $u \in \{-p,\ldots,p\}$. These equations can be rewritten in terms of time domain quantities. For this, let $\hat{\boldsymbol{\Gamma}}(u)$ be the empirical covariance function, which is related to the periodogram by

$$\hat{\boldsymbol{\Gamma}}(u) = \int_\Pi \mathbf{I}^{(T)}(\lambda) e^{\mathrm{i}\lambda u} d\lambda. \quad (11)$$

Noting that similarly $\boldsymbol{\Gamma}_{\boldsymbol{\theta}}(u) = \int_\Pi \mathbf{f}_{\boldsymbol{\theta}}(\lambda) e^{\mathrm{i}\lambda u} d\lambda$, we obtain the following likelihood equations.

**Theorem 4.1.** *Let $\hat{\boldsymbol{\theta}}_T$ be the Whittle estimator of the parameter $\boldsymbol{\theta}_0$ in the graphical interaction model $GI(p,G)$. Then $\hat{\boldsymbol{\theta}}_T$ is a solution of the following equation system:*

*(i) for $a,b \in V$ with $a=b$ or $a \mathrm{\:\!-\!\:} b \in E$*

$$\Gamma_{ab,\hat{\boldsymbol{\theta}}_T}(u) = \hat{\Gamma}_{ab}(u), \quad u=-p,\ldots,p;$$

*(ii) for $a,b \in V$ with $a \neq b$ and $a \mathrm{\:\!-\!\:} b \notin E$*

$$\Gamma^{(i)}_{ab,\hat{\boldsymbol{\theta}}_T}(u) = 0, \quad u=-p,\ldots,p.$$

These equations are similar to the likelihood equations in ordinary Gaussian graphical models (cf Lauritzen 1996). In fact, together with the condition that $\boldsymbol{\Gamma}^{(i)}_{\hat{\boldsymbol{\theta}}_T}(u) = 0$ for $|u| > p$, these equations can be seen as the likelihood equations obtained in a Gaussian graphical model for the random variables $X_v(t)$ with $v \in V$ and $t \in \mathbb{Z}$, where the graph is specified by the zero constraints on $\boldsymbol{\Gamma}^{(i)}$ and $\hat{\boldsymbol{\Gamma}} = \big(\hat{\boldsymbol{\Gamma}}(u-v)\big)_{u,v \in \mathbb{Z}}$ is the empirical covariance matrix of the variables. This is not surprising by the way the Whittle likelihood approximates the log-likelihood function in (8): Whittle's approximation basically neglects the edge effects due to observing only a finite horizon by substituting asymptotic approximations for the finite sample parameters $\det \boldsymbol{\Gamma}_{\boldsymbol{\theta},T}$ and $\boldsymbol{\Gamma}^{-1}_{\boldsymbol{\theta},T}$.

The asymptotic properties of the Whittle estimator are well known (e.g., Dzhaparidze and Yaglom 1983). For the formulation of the asymptotic distribution, let $\boldsymbol{\Xi}(\boldsymbol{\theta}) = (\xi_{ij}(\boldsymbol{\theta}))$ be the matrix with entries

$$\xi_{ij}(\boldsymbol{\theta}) = \frac{1}{4\pi} \int_\Pi \text{tr}\Big[\mathbf{f}_{\boldsymbol{\theta}}(\lambda) \frac{\partial \mathbf{f}_{\boldsymbol{\theta}}(\lambda)^{-1}}{\partial \theta_i} \mathbf{f}_{\boldsymbol{\theta}}(\lambda) \frac{\partial \mathbf{f}_{\boldsymbol{\theta}}(\lambda)^{-1}}{\partial \theta_j}\Big] d\lambda.$$

Furthermore let $\mathbf{P}_G$ be a projector matrix such that $\mathbf{P}_G \boldsymbol{\theta}$ is the vector of unconstrained parameters in GI($p,G$). Then, if $\boldsymbol{X}_V$ belongs to the GI($p,G$) model with parameter $\boldsymbol{\theta}_0$, the Whittle estimator $\hat{\boldsymbol{\theta}}_T$ is asymptotically normally distributed,

$$\sqrt{T}\left(\hat{\boldsymbol{\theta}}_T - \boldsymbol{\theta}_0\right) \xrightarrow{\mathcal{D}} \mathcal{N}\big(\mathbf{0}, \boldsymbol{\Lambda}(\boldsymbol{\theta}_0)\big),$$

where $\boldsymbol{\Lambda}(\boldsymbol{\theta}_0) = \mathbf{P}'_G \big(\mathbf{P}_G \boldsymbol{\Xi}(\boldsymbol{\theta}_0) \mathbf{P}'_G\big)^{-1} \mathbf{P}_G$.

We note that, although the Whittle estimator $\hat{\boldsymbol{\theta}}_T$ is consistent for $\boldsymbol{\theta}_0$, the likelihood equations in general may have multiple solutions. Thus, a solution of the likelihood equations may be only a local minimum of the Whittle log-likelihood $\ell_T^w(\boldsymbol{\theta})$.

## 5 ITERATIVE ESTIMATION

For Gaussian graphical models, there exist two iterative algorithms for solving the likelihood equations: iterative proportional scaling (e.g., Lauritzen 1996) and the algorithm by Wermuth and Scheidt (1977). The former requires that all constraints are preserved throughout the iterations and is hard to realize in the present situation. We therefore discuss an adapted version of the second algorithm.

Let $C_{p,G}$ be the set of all $(a,b,u) \in V \times V \times \mathbb{Z}$ such that $a$ and $b$ are distinct and not adjacent in $G$ or $|u| > p$. Then $C_{p,G}$ represents the entries in the infinite-dimensional inverse covariance matrix $\boldsymbol{\Gamma}^{(i)}$ that are constrained to zero in the GI($p,G$) model, that is, we have in the GI($p,G$) model

$$(a,b,u) \in C_{p,G} \Rightarrow \Gamma^{(i)}_{ab}(u) = 0.$$

For the application of the algorithm by Wermuth and Scheidt (1977), which preserves only part of the constraints in each iteration, we define subsets $C_i$ specifying the constraints to be preserved alternately. To this end, let $\{a_i, b_i\}$, $i = 1, \ldots, m$, be an enumeration of all distinct vertices $a_i, b_i$ that are not adjacent in $G$. Then we set

$$C_0 = \big\{(a,b,u) \in V \times V \times \mathbb{Z} \big| |u| > p\big\}$$

and, for $i = 1, \ldots, m$,

$$C_i = \big\{(a,b,u) \in V \times V \times \mathbb{Z} \big| \{a,b\} = \{a_i, b_i\}\big\}.$$

Here, the set $C_0$ represents the constraints imposed by the order $p$ of the GI($p,G$) model while the sets $C_i$ indicate the constraints due to the absence of the edges $a_i$ — $b_i$ in $G$. Furthermore, we have $C_{p,G} = \cup_{i=0}^m C_i$.

Starting with $\hat{\boldsymbol{\Gamma}}_0(u) = \hat{\boldsymbol{\Gamma}}(u)$ for all $u \in \mathbb{Z}$, where $\hat{\boldsymbol{\Gamma}}(u)$ is the empirical covariance function given by (11), the algorithm proceeds by solving in the $n$-th step the equation system

$$\hat{\Gamma}_{ab,n}(u) = \hat{\Gamma}_{ab,n-1}(u) \quad \forall (a,b,u) \notin C_{n \bmod m+1},$$
$$\hat{\Gamma}^{(i)}_{ab,n}(u) = 0 \quad \forall (a,b,u) \in C_{n \bmod m+1}.$$

In order to avoid working with infinite-dimensional matrices, the computations are carried out in the frequency domain:

For $n \bmod m + 1 = 0$, the constraints in $C_0$ define a VAR($p$) model without further constraints. Thus, the solution of the above equation system satisfies the Yule-Walker equations

$$\hat{\boldsymbol{\Gamma}}_{n-1}(u) = \sum_{v=1}^p \hat{\boldsymbol{\Gamma}}_{n-1}(u-v) \mathbf{a}(v)' + \boldsymbol{\Sigma} \delta_{u0},$$

which are solved by the Yule-Walker estimates $\hat{\mathbf{a}}_n(1), \ldots, \hat{\mathbf{a}}_n(p)$ and $\hat{\boldsymbol{\Sigma}}_n$ (e.g., Brockwell and Davis 1991). From these, the spectral density can be obtained by

$$\hat{\mathbf{f}}_n(\lambda) = \hat{\mathbf{A}}_n(e^{-i\lambda})^{-1} \hat{\boldsymbol{\Sigma}}_n \hat{\mathbf{A}}_n(e^{i\lambda})'^{-1},$$

where $\hat{\mathbf{A}}_n(z) = \mathbf{I} - \hat{\mathbf{a}}_n(1)z - \ldots - \hat{\mathbf{a}}_n(p)z^p$.

For $n \bmod m + 1 = i > 0$, the constraints in $C_i$ are equivalent to $f^{-1}_{a_i b_i}(\lambda) = 0$. Consequently, the iteration step can be formulated as

$$\hat{f}_{ab,n}(\lambda) = \hat{f}_{ab,n-1}(\lambda) \quad \forall \lambda \in [-\pi,\pi] \quad \text{if } \{a,b\} \neq \{a_i,b_i\},$$
$$\hat{f}^{-1}_{ab,n}(\lambda) = 0 \quad \forall \lambda \in [-\pi,\pi] \quad \text{if } \{a,b\} = \{a_i,b_i\}.$$

This can be accomplished by setting

$$\hat{f}_{a_i b_i,n}(\lambda) = \hat{\mathbf{f}}_{a_i S_i, n-1}(\lambda) \hat{\mathbf{f}}_{S_i S_i, n-1}(\lambda)^{-1} \hat{\mathbf{f}}_{S_i b_i, n-1}(\lambda)$$

for $\lambda \in [-\pi, \pi]$, where $S_i = V \setminus \{a_i, b_i\}$ (cf Wermuth and Scheidt 1977).

Speed and Kiiveri (1986) have shown the convergence of the algorithm to a solution of the likelihood equations in the case of Gaussian graphical models; the proof for the present situation is similar. After the $n$-th step, an estimate $\hat{\boldsymbol{\theta}}_T$ for $\boldsymbol{\theta}$ can be obtained by inverting the spectral matrix $\hat{\mathbf{f}}_n(\lambda)$ and computing the inverse covariances $\hat{\boldsymbol{\Gamma}}^{(i)}_n(u)$ by (4). Alternatively, if the algorithm is stopped at $n \bmod m + 1 = 0$, it also provides the estimates in the parameterization of the VAR($p,G$) model.

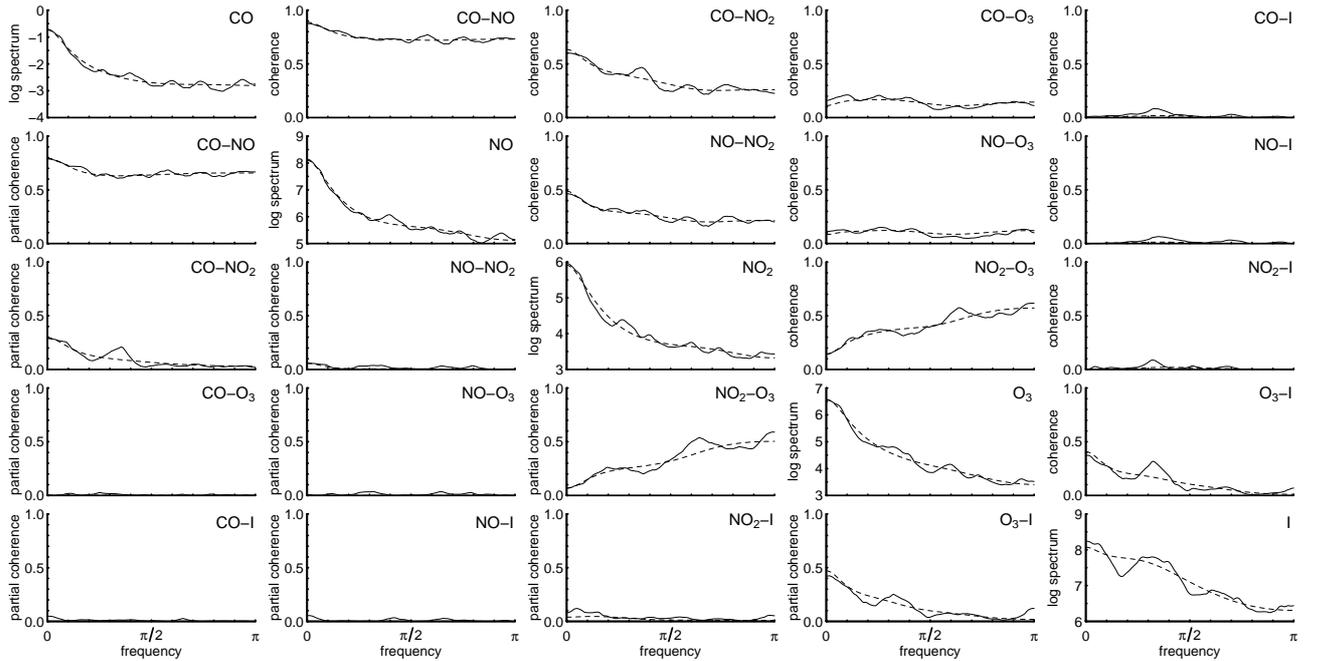

Figure 2: Estimated spectral densities, coherencies, and partial coherences for air pollution data: nonparametric estimates (*solid lines*) and parametric fit (*dashed lines*) obtained by minimizing the BIC criterion.

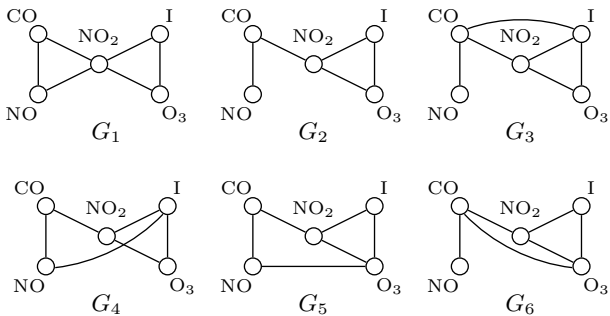

Figure 1: Undirected graphs with lowest BIC value for the air pollution data.

Table 1: Results of model selection for the air pollution data: graphs and orders $p$ with the lowest BIC values.

| Graph | BIC ($p=3$) | BIC ($p=4$) |
|---|---|---|
| $G_1$ | 686.0 | 705.6 |
| $G_2$ | 687.7 | 687.0 |
| $G_3$ | 708.8 | 722.1 |
| $G_4$ | 714.9 | 733.1 |
| $G_5$ | 716.1 | 733.5 |
| $G_6$ | 718.7 | 748.1 |

## 6 EXAMPLE

We illustrate the proposed graphical modelling approach by application to a five-dimensional time series that has been analysed previously by the nonparametric frequency domain approach in Dahlhaus (2000). The time series consists of $T = 4386$ measurements of the concentrations of four air pollutants—carbon monoxide (CO), nitrate monoxide (NO), nitrate dioxide ($NO_2$), and ozone ($O_3$)—and the global radiation intensity (I) recorded from January 1991 to December 1992 in Heidelberg (6 equidistant recording per day). For more details, we refer to Dahlhaus (2000).

In order to learn the conditional independence graph of $\boldsymbol{X}_V$ from the data, we fitted graphical interaction models of orders ranging from $p = 1$ to $p = 6$ for all possible undirected graphs. For each fitted model, the parameters were computed in their autoregressive form, that is, as $\hat{\mathbf{a}}_{p,G}$ and $\hat{\boldsymbol{\Sigma}}_{p,G}$. From these, we obtained the BIC score (Schwarz 1978)

$$\text{BIC}(p,G) = T \det \hat{\boldsymbol{\Sigma}}_{p,G} + \log(T)\, q_{p,G},$$

where $q_{p,G}$ is the number of parameters in the GI(p,G) model. The BIC criterion was minimized for order $p = 3$ and the graph $G_1$ in Figure 1. Figure 2 shows the parametric estimates of the spectral densities and the spectral and partial spectral coherences for the selected model (*dashed lines*); these fit the nonparametric estimates (*solid lines*) reasonably well. Notice that in the four lower left plots the parametric estimates for the partial spectral coherence are identical to zero as required by (1).

Table 1 shows the BIC scores for the six best models of orders 3 and 4 (models with different orders led to higher BIC scores). Here, the scores for the best model GI(3,$G_1$) and the models GI($p$,$G_2$) with $p = 3, 4$ differ by less than 2. This indicates that there is not enough evidence in the data to discriminate between the best model and the competing models GI($p$,$G_2$) with $p = 3, 4$ (e.g., Raftery 1995). We note that the graph $G_2$ was also selected by the nonparametric approach in Dahlhaus (2000).

# 7 DISCUSSION

In this paper, we have presented a parametric approach for graphical interaction modelling of stationary Gaussian processes. Graphical interaction models provide a simple description of the dependence structure of stationary processes by undirected graphs. We have proposed a new parametrization of vector autoregressive models in terms of inverse covariances. With this parametrization, the conditional independence restrictions encoded by an undirected graph lead to simple zero constraints on the parameters.

We have discussed parameter estimation based on Whittle's approximation to the log-likelihood function and proposed to compute the resulting estimates by an adapted version of the iterative algorithm by Wermuth and Scheidt (1977). This algorithm not only has clear convergence properties, but also provides estimates alternatively in the parameterization of the graphical interaction model GI($p$,$G$) and of the graphical vector autoregressive model VAR($p$,$G$).

**Acknowledgements**